\documentstyle[aps,twocolumn,psfig]{revtex}
\newcommand{\be}{\begin{equation}} 
\newcommand{\ee}{\end{equation}}
\newcommand{\bea}{\begin{eqnarray}}
\newcommand{\eea}{\end{eqnarray}}
\begin{document}
\draft
\title{ 
\begin{flushright}
SINP/TNP/97-22 \\
%$\langle$nucl-th/97120xx$\rangle$\\
\end{flushright}
\bf {Effect of Colour Singletness of Quark-Gluon Plasma in 
Quark-Hadron Phase Transition}} 
\author
{ Munshi Golam Mustafa$^{a}$, Dinesh Kumar Srivastava$^{b}$
and Bikash Sinha$^{a,b}$}
\medskip
\address
{ $^a$ Saha Institute of Nuclear Physics, 1/AF Bidhan Nagar, 
Calcutta-700 064, India.}
\medskip
\address
{ $^b$ Variable Energy Cyclotron Centre, 1/AF Bidhan Nagar, 
Calcutta 700 064, India.}
\medskip
\date{\today}
\maketitle

\begin{abstract}
\noindent Consequences of the constraint of SU(3) colour singletness 
of quark-gluon plasma are studied. This restriction increases
the free energy barrier for the formation of hadronic bubble 
in supercooled phase and influences significantly the dynamics of the 
initial stage of quark-hadron phase 
transition. It also introduces terms dependent on the volume occupied 
by the plasma in the energy density and the pressure. These modifactions
vanish in the limit of an infinite volume. The last stage of
the hadronization of the QGP likely to be formed in relativistic
heavy ion collisions is necessarily characterized by a decreasing
volume containing the quark matter, and thus these corrections
become important. The nucleation of plasma droplets at AGS energies is
also seen to be strongly affected by the requirement of colour
singletness, and the choice of prefactor.
\end{abstract}

{}
\narrowtext

%\newpage
\section{{\bf Introduction}}
\label{sec: int}
The success of the quark model,  the quantum
chromodynamics (QCD), and the non-observability of the free partons
($q$, ${\bar q}$, $g$) has entailed the concept of confinement. QCD, 
the theory of strong interactions, is not perturbative at large
distances. Thus, the confinement itself can not be treated
perturbatively. 
There are reasons to believe that the confinement of partons
inside hadrons may not survive  collisions between heavy nuclei at 
relativistic energies. In such collisions, the two nuclei masquerading as 
clouds of space and time like partons
pass through each other, leaving behind a high density plasma of 
quarks, antiquarks, and gluons (QGP)~\cite{klaus} in their wake, 
in the region between the two receding fronts of the leading particles.
This plasma expands and cools, the energy density becomes
low enough and a phase transition to a hadron gas takes place around
the critical temperature, $T_C$.
A dynamic treatment  of this phase transition is a problem of considerable
interest.

An understanding of the QCD phase transition requires a knowledge of the 
equation of state as well as the kinetics of phase transition.
If QCD has a first order phase transition, it may proceed with a
supercooling of the QGP followed by a nucleation and growth of
hadronic bubbles~\cite{joe,joe1}, releasing the latent heat as the phase
transition progresses. In a superheated hadronic matter, on the other hand, 
nucleation of a QGP droplet may also proceed similarily.

For a first order phase transition the rate for hadronic bubble/plasma droplet 
nucleation can be estimated in the frame work of homogeneous nucleation
theory~\cite{ll}
\be
I=I_0 \exp (-\triangle F^\star/T) \ \ , \label{nra}
\ee
where $I_0$, which has the dimensions of 1/fm$^4$-is called the prefactor, 
$T$ is the temperature, and $\triangle F^\star$ is the change in free
energy accompanying the formation of a critical size hadronic bubble/plasma
droplet. 
These  dimensional arguments were used in a large number of
studies in the past to replace $I_0$ with  $T^4$ or $T_C^4$.
This unsatisfactory state of affairs was corrected recently by 
Kapusta and Csernai~\cite{joe2}.
They computed the dynamical prefactor in a course-grained effective
field theory approximation to QCD. This dynamical factor 
influences the growth rate and statistical fluctuations and also accounts
for the available phase space. 
 
Csernai et al.~\cite{joe,joe1} have also used a nucleation
rate equation with this realistic dynamical prefactor to study the time
evolution of expanding QGP as it converts to hadronic matter. 
They noted a substantial deviation
from an idealized Maxwell construction  that has often been employed
as a  model of hadronization~\cite{max}. 
Obviously, such an idealized phase transition assumes a 
QCD nucleation rate which is much larger than the rate of expansion. This is 
not necessarily true.

In all such studies, QGP is generally described as an ideal gas 
of quarks, antiquarks and gluons, essentially  
described by the Stefan-Boltzmann law. Lattice calculations~\cite{det} 
have provided ample evidence that even at fairly
high temperatures, colour singlet objects like multi-quark cluster
($q{\bar q}$, $qqq$, ${\bar q}{\bar q}{\bar q}$, $\cdots$) propagate
in the plasma. One may account for this `interaction' by  requiring  
that all physical states be colour singlet with respect to the SU(3) 
colour gauge group~\cite{red,aub,go,mus}.

It has recently been shown~\cite{mus1} 
that restricting the quark partition function
to be colour singlet of SU(3) colour gauge group  amounts to
reordering the thermodynamic potential in terms of the
colourless multi-quark modes ($q{\bar q}$, $qqq$, 
${\bar q}{\bar q}{\bar q}$, $\cdots$) at any given temperature.
Under a suitable confining mechanism, these could evolve into 
colour singlet hadrons/baryons at low temperatures. This is 
also in accord with the $``$preconfinement" property of QCD noted by 
Amati and Veneziano~\cite{ave} quite sometime ago where the
cascading and fragmenting partons produced in hadronic collisions 
rearrange themselves into colour singlet clusters which ultimately
evolve into hadrons~\cite{mw,eg}. These considerations convince us 
that it is important to incorporate the {\it dynamic} requirement of
colour singletness of the quark-matter which $``$tunnels" into hadronic 
matter phase space~\cite{eg}.
%It is also known that the coloursingletness affects 
%the quark number susceptibility significantly, which is an order
%parameter for chiral phase transition. 

In the present work we study the consequences of the
incorporation of colour singlet equation of state for plasma 
on the dynamics of quark-hadron phase transition. 

\section{{\bf Equation of State}}
\label{eeq}
\subsection{{\bf Colour Singlet Equation of State
for a Quark-Gluon Plasma}}
\label{qeq}
Consider a quark-gluon plasma  consisting of `u' and `d' quarks, and
gluons. The grand canonical partition function~\cite{aub,go,mus} 
subject to colour singletness can be written as

\be
{\cal Z}(\beta, V_q) = {{\mathrm {Tr}}} \ \left ( {\hat {\cal P}}
\ e^{-\beta \hat H} \right ) \ , \label{pf}
\ee

\noindent where $\beta \ = \ {1/T}$ is the inverse temperature,
$V_q$ is the volume, $\hat H$ is the Hamiltonian of the physical
system, and ${\hat {\cal P}}$ is the colour  projection operator.  
For a baryon free plasma, this can be simplified after a 
considerable amount of group theoretic algebraic 
manipulations~\cite{aub,mus} to give,

\be
{\cal Z}(\beta, V_q)   =   {\sqrt 3\over
{3\pi}} \ \Big [ \ {8V_q\over {3 \beta^3}} \Big ]^{-4} 
\, \exp \ \Big [ \ {a_q V_q\over
{\beta^3}}  \ \Big ]
\ ,\label{fpf}
\ee
where $a_q=37\pi^2/90$.
Now the free energy of the baryon-free colourless 
quark-gluon gas is obtained as, 
\be
F_q \ = \ - \ T \ \ln {\cal Z}(T,V_q) \ + BV_q \ . \label{fq}
\ee
One may now write for the energy density
\bea
e_q\ &=& \ E_q\over{V_q} \nonumber\\
  &=& \ {T^2\over{V_q}}{\partial\over{\partial T}}
\left [\ln {\cal Z}(\beta,V_q) \right ] \ + \ B \nonumber \\
& = & \ B \ + \ 3a_q T^4 \ -{ 12 T\over{V_q}} \ , \label{eq}
\eea

The pressure of the above quark-gluon system is given as
\bea
P_q \ &=& \ - \left ({\partial F\over
{\partial V}}\right )_T  \nonumber \\
& =&  - \ B \ + \ a_q T^4 \
- \ {4T\over {V_q}} \ . \label{pq}
\eea

We note that the colour singletness introduces  corrections to the
normally assumed expresssions for the energy density  and
the pressure which vanish for an infinite volume. Now consider the hadronic 
phase arising as a result of first order quark-gluon/hadron transition
through the nucleation of hadronic bubble in the bulk QGP.
With the passage of time, more and more of the 
quark matter will get converted to hadronic matter. The later stage of 
the process of hadronization will be characterized by a decreasing volume 
occupied by the quark matter. If we believe the above equations of state,
the volume occupied by the plasma can not be vanishingly small.
The colour singletness will again have important consequences for 
dynamics of the phase transition during the later stage of hadronization. 

Here, we would also like to point out that the colour singletness has 
important bearing on the nucleation of a hadronic bubble in the plasma
which we will see in sec. III.

\subsection{{\bf Equation of State for Hadron Gas}}
\label{heq}
We model the hadronic phase as  a gas of massless pions.
The energy density and the pressure of such a system can be
written as
\be
e_h=3a_hT^4 \ , \label{fh}
\ee
\noindent and
\be
P_h=a_hT^4 \ , \label{ph}
\ee
where $a_h=\pi^2/30$.
\section{{\bf Supercooling and Nucleation}}
\label{nhb}
Nucleation in a pure phase like QGP proceeds via creation of a
hadronic bubble due to statistical fluctuations in a supercooled plasma.
The bubble is made up of hot pion gas and is surrounded by
colour singlet plasma of volume, $V_q=(V-V_b)$, where $V$ is 
initial volume of plasma. $V_b=4\pi {R_b}^3/3$ represents an 
excluded volume corresponding to hadronic bubble. One can think of
curving out a colour singlet piece of palsma and replacing
it with a hadronic bubble. The fields in
plasma obey the bag boundary conditions, staying outside the hadronic
bubble~\cite{mar,bura}.  If the radius of the bubble is $R_b$,
the change in free energy can be written~\cite{mus,mar} within the bag 
model as,
\bea 
\triangle F & =& T\ln \left ( \pi \sqrt 3\right ) + 4T\ln \left ( 
{8\over 3} V_b T^3 \right ) + a_q V_bT^4 \nonumber \\
&-& (B +P_h)V_b +4\pi {R_b}^2 \sigma \ , \label{df}
\eea
\noindent where $P_h$ is the pressure of the hadron gas
given in Eq.(\ref{ph}) and $\sigma$ is the surface free energy of 
the quark-gluon/hadron interface. The first 
two terms in Eq.(\ref{df}) are due to SU(3) colour singlet restriction.
They increase the barrier for $\triangle F$ required to form hadronic
bubble in plasma (Fig.1). We shall see later that it has an
important effect during the initial stage of QCD phase transition.

Recall that one can derive the critical radius (${R_b}^\star$) of
hadronic bubble by minimizing the change in free energy, $\triangle F$,
with respect to $R_b$. If they are too small ($R < {R_b}^\star$), they 
will shrink and vanish. If they are large ($R > {R_b}^\star$), they 
will grow. Now, minimizing $\triangle F$ with respect to $R_b$, one 
gets
\be
{12 T\over {R_b}^\star} -4\pi (B-a_{qh}T^4){R_b}^{\star 2} +
8\pi\sigma {R_b}^\star = 0 \ , \label{ddf}
\ee
which will yield the critical radius of hadronic bubble ${R_b}^\star$.
We have further defined, $a_{qh}=a_q-a_h$.
If the first term in Eq.(\ref{ddf}), which has its origin in the
requirement of colour singletness is neglected, one obtains the critical 
radius, ${R_b}^\star={2\sigma/(B-a_{qh}T^4)}$, for a nonsinglet case.
Making a substitution 
$x=1/{R_b}^\star$, Eq.(\ref{ddf}) can be written as
\be
x^3 + ax -b = 0 \ \ , \label{sdf}
\ee
where $a=2\pi\sigma/3T$ and $b=\pi(B-a_{qh}T^4)/3T$. The physical 
solution of Eq.(\ref{sdf}) gives the critical radius of hadronic 
bubble for the colour singlet case, as,
\be
{R_b}^{\star }= 3z/\left (3z^2-a^2\right ) ; \ \ \ z^3=
\left [ {b/2} + \sqrt {{a^3/27} +{b^2/4}} \right ]
\ \ . \label{xrb}
\ee

Now, the change in free energy for the creation of a hadronic bubble 
having the critical radius ${R_b}^\star$ is
\be
\triangle F_\star = \triangle F |_{R={R_b}^\star} \ . \label{dfs}
\ee

The rate for the nucleation of hadronic phase out of the plasma
phase is usually estimated by Eq.(\ref{nra}) with prefactor, $I_0$.
As remarked earlier the prefactor, $I_0$,  
has been  calculated by Csernai and Kapusta~\cite{joe2}
in an effective field theory approximation to QCD as
\be
I_0 \ = \ {16\over {3\pi}} \left ( {\sigma\over {3T}} \right )^{3/2}
{\sigma \eta_q {R_b}^\star \over {\xi_q^4 (\triangle w)^2}} \ \ .
\ee
\noindent Here, $\eta_q$ and $\xi_q$ are, respectively, the shear
viscosity and correlation length in the plasma phase, and
$\triangle w$ is the difference in the enthalpy densities of
the two phases. We use the same parameter set as used in 
Refs.~\cite{joe,joe1},
e.g., $B^{1/4}=235$ MeV, $\xi_q=0.7$ fm, $\eta_q=14.4T^3$ and
$T_C=169$ MeV. Next we closely follow the arguments of 
Refs.~\cite{joe,joe1} to obtain the dynamics of the phase transition.

Once the nucleation rate is known, one can calculate the
(volume) fraction of space $h(\tau)$ converted to hadronic gas
at a proper time $\tau$. This proper time is measured in
the local comoving frame of an expanding system. For this purpose
one needs a kinetic equation which involves $I$ as the source. If
the system cools to $T_C$ at a time ${\tau}_C$, then at some later
time $\tau$ the fraction of the space which has converted to
hadronic gas is~\cite{joe,joe1}
\be
h(\tau) \ = \ \int_{{\tau}_C}^\tau {\mathrm d}\tau^\prime
I\left ( T(\tau^\prime)\right ) \left [ 1-h(\tau^\prime ) \right ]
V(\tau^\prime,\tau) \ \ , \label{hf}
\ee
\noindent where, $V(\tau^\prime,\tau)$ is the volume of a bubble
at a time $\tau$ which has nucleated at an  earlier time $\tau^\prime$.
This also takes into account bubble growth. How rapidly does the
bubble grow after nucleation?
Usually a critical size bubble is metastable and will not grow without
a perturbation. Pantano and Miller~\cite{mp} have numerically 
computed the growth of bubbles using a relativistic hydrodynamics. 
The asymptotic radial growth velocity was found to be consistent with 
the growth law
\be
v(T) \ = \ v_0 \left ( 1- T/T_C \right )^{3/2} \ \ , \label{vg}
\ee
\noindent where $v_0$ is a model-dependent constant.
We shall use $v_0=3c$ as it has been argued~\cite{joe,joe1}
that the Eq.(\ref{vg}) is intended to be applied
only as long as $T > 2T_C/3$ so that the growth velocity stays below
the speed of sound of a massless gas, $c/\sqrt{3}$.
Thus  the growth of the bubble can be approximated as,
\be
V(\tau^\prime,\tau) \ = \ {4\over 3}\pi \left [ {R_b}^\star \left (
T(\tau^\prime) \right ) \ + \ {\int_{{\tau}^\prime}}^\tau
{\mathrm d}\tau^{\prime \prime} v\left (T(\tau^{\prime \prime})\right )
\right ]^3 \ \ . \label{ve}
\ee

Now one needs a dynamical equation which couples the time
evolution of the temperature to the fraction of space
converted to hadronic gas. For this purpose we use both
Bjorken longitudinal hydrodynamics and 
Cooper-Frye-Sch${\ddot{\mathrm o}}$nberg
spherical hydrodynamics as Refs.~\cite{joe,joe1}. In Bjorken
model~\cite{bj} the time evolution of energy density $e$ is given as
\be
{{\mathrm d}e\over{{\mathrm d}\tau}} \ = \ -{w\over \tau} \ 
\ \ ,\label{bh}
\ee
whereas in Cooper-Frye-Sch${\ddot{\mathrm o}}$nberg model~\cite{cf} it is
\be
{{\mathrm d}e\over{{\mathrm d}\tau}} \ = \ -{3w\over \tau} \ 
\ \ .\label{cfs}
\ee
In Eq.(\ref{cfs}) the factor of 3 appears because of the spatial 
expansion along three dimensions rather than one dimension. In order for
an easy comparison we consider initial conditions as in
Refs.~\cite{joe,joe1} which are likely to be achieved in collisions
involving two gold nuclei at RHIC energies.
Thus for longitudinal expansion we take $\tau_i = 3/8$ fm/c and $T_i=2T_C$.
The temperature decreases as $T(\tau) \propto \tau^{-1/3}$
until the time $\tau_C = 3$ fm/c  when the temperature becomes $T_C$.
In the case of spherical expansion  we consider $T_i=2T_C$ and
$\tau_i={\sqrt 2} R_{\mathrm{nucl}}$ and the temperature decreases
like $T(\tau) \propto \tau^{-1}$ until the time $\tau_C =18$ fm/c 
corresponding to half density radius of a gold nucleus (For details 
see Refs.~\cite{joe,joe1}). The Eqs.(\ref{bh}) and (\ref{cfs}) are
essentially the statement of energy conservation in the
respective picture which assume kinetic equilibrium but not phase
equilibrium. The energy density can be written~\cite{joe,joe1} as 
\be
e(T) \ = \ h(\tau)e_h(T) \ + \ \left [ 1-h(\tau)\right ] e_q(T) \ \ .
\label{ec}
\ee
Here, $e_h$ and $e_q$ are the the energy densities in hadronic
and plasma phases at temperature $T$, and similarly for $w$.

\section{Dynamics of Phase Transition: Longitudinal and
Spherical Expansion}
\label{dyns}

In Fig.1 the free energy difference, $\triangle F$ in Eq.(\ref{df}) 
as a function of hadronic bubble radius for a fixed bag constant
($B^{1/4}=235$ MeV), surface tension ($\sigma=50$ MeV/fm$^2$) and  
temperature ($T=160$ MeV) is displayed. Results are given with and 
without colour singlet requirement of the system. One
finds a significant increase in $\triangle F$ if the constraint of 
colour singletness is imposed. We will see below that it has
important consequences on the time evolution of expanding QGP as it
converts to hadronic matter through nucleation of hadronic bubble.

%%%%%%%%%%%%%%%%%%%%%%%% begin of Fig.1 %%%%%%%%%%%%%%%%%%%%%%%%%%%
\vskip 0.3in
\begin{figure}
\centerline{
\psfig{figure=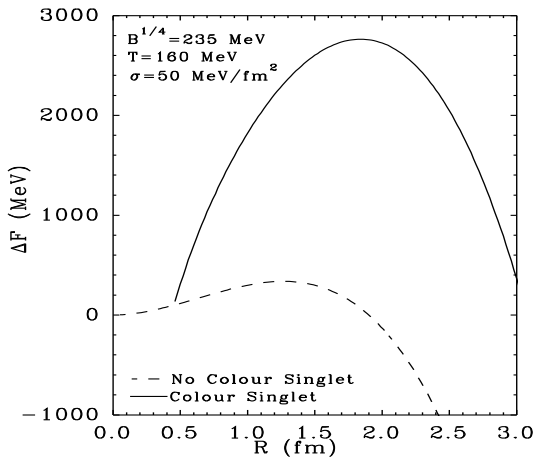,height=6cm,width=7cm}}
\vskip 0.3in
\caption{ The free energy difference $\triangle F(R)$ for creation of a
hadronic bubble in quark-gluon plasma.}
\end{figure}
\vskip 0.3in
%%%%%%%%%%%%%%%%%%%%%%%%end of Fig.1 %%%%%%%%%%%%%%%%%%%%%%%%%%%

In Fig.2 we show the variation of the temperature with proper time as 
the matter undergoes longitudinal expansion with the initial
conditions given in sec. 3. We have also given the
results for an adiabatic phase transition for a comparision.
The value of $\sigma$ used here is 50 MeV/fm$^2$. The
general features are similar to those of Refs.~\cite{joe,joe1}.
%%%%%%%%%%%%%%%%%%%%%%%% begin of Fig.2 %%%%%%%%%%%%%%%%%%%%%%%%%%%
\vskip 0.3in
\begin{figure}
\centerline{
\psfig{figure=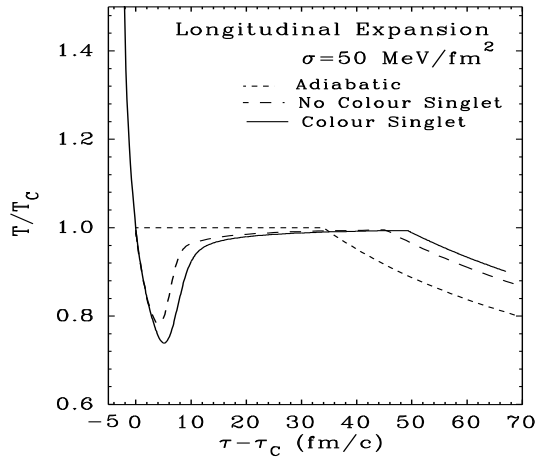,height=6cm,width=7cm}}
\vskip 0.3in
\caption{The temperature as function of proper time for a hadronizing
quark-gluon plasma in a central high-energy nucleus-nucleus collision 
for matter undergoing longitudinal expansion. The initial conditions 
correspond to QGP formation at  BNL - RHIC energies.}
\end{figure}
\vskip 0.3in
%%%%%%%%%%%%%%%%%%%%%%%%end of Fig.2 %%%%%%%%%%%%%%%%%%%%%%%%%%%
 The matter
continues to cool below $T_C$ until nucleation of hadronic bubble sets
in. For colour singlet case the degree of supercooling is about 30$\%$,
$i.e.$, 12.5$\%$ more than that for the nonsinglet case. Once nucleation and
growth of bubble start, the system reheats near $T_C$ due to release of
latent as the phase transition progresses. When temperature approaches
$\sim$0.95$T_C$ nucleation of further bubble formation ceases off and
the transition proceeds because of growth of previously nucleated
bubbles. However, the system can not reheat upto $T_C$ because bubble
growth (Eq.(\ref{ve})) reduces to zero as $T_C$ is approached. Recall
that nucleation in colour singlet case will be delayed due to increase in
height of $\triangle F$ which in turn also slows down the phase transition.
This results in $10\%$ extra entropy generation in the processs as compared 
to colour nonsinglet case.

Fig.3 shows the variation of critical radius as a function of proper 
time for
hadronic bubble undergoing longitudinal expansion with and without
colour singlet requirement. It clearly indicates that the early stage
of nucleation is characterized by a much larger critical size of
nucleated hadronic bubble ($\sim 1$ fm) for colour singlet case as 
compared to $\sim 0.5$ fm for colour nonsinglet case. At later 
times ($\geq$ 12 fm/c) the critical size of nucleated hadronic bubble
remains lower until the phase transition is completed if the colour
singlet requirement is implemented in the system.  In these initial
studies we have not included the modification due to bubble fusion which
should be of interest.

%%%%%%%%%%%%%%%%%%%%%%%% begin of Fig.3 %%%%%%%%%%%%%%%%%%%%%%%%%%%
\vskip 0.3in
\begin{figure}
\centerline{
\psfig{figure=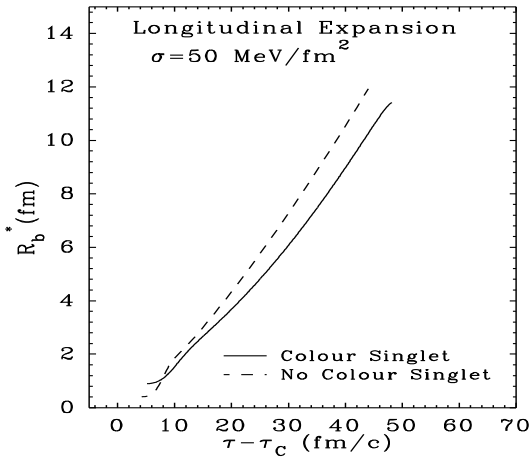,height=6cm,width=7cm}}
\vskip 0.3in
\caption{Radius of critical hadronic bubbles as a function of time, in a
hadronizing quark-gluon plasma.}
\end{figure}
\vskip 0.3in
%%%%%%%%%%%%%%%%%%%%%%%%end of Fig.3 %%%%%%%%%%%%%%%%%%%%%%%%%%%

%%%%%%%%%%%%%%%%%%%%%%%% begin of Fig.4 %%%%%%%%%%%%%%%%%%%%%%%%%%%
\begin{figure}
\centerline{
\psfig{figure=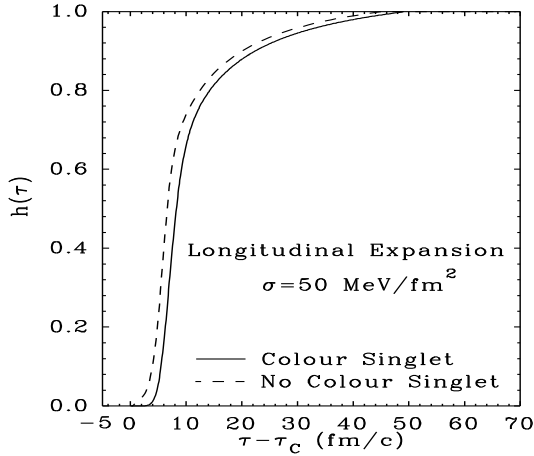,height=6cm,width=7cm}}
\vskip 0.3in
\caption{ The volume-fraction of space occupied by the hadronic matter 
as a function of time, in a hadronizing quark-gluon plasma.}
\end{figure}
\vskip 0.3in
%%%%%%%%%%%%%%%%%%%%%%%%end of Fig.4 %%%%%%%%%%%%%%%%%%%%%%%%%%%
In Fig.4 we show the volume fraction converted to hadronic matter
undergoing longitudinal expansion as a function of proper time. 
The delay in completion of QCD phase transition due to colour singletness
is seen clearly. As the phase transition progresses via nucleation 
of hadronic bubble, the available space  will be progressively occupied 
by hadronic matter. We would like to make an amusing observation here. If
we believe the equations of state for the QGP as discussed earlier
(Eqs.(\ref{eq}) and (\ref{pq})), the
volume occupied by plasma can {\it not} be vanishingly small. This necessarily
implies a remnant of quark matter when the process of hadronization is
over. We find that a fraction of quark matter of mass $\sim$ (5-10) GeV 
having volume $\sim$ 10--15  fm$^3$ remains unconverted at the end of the 
hadronization in the central region, per unit rapidity. One can have 
interesting speculations about such a remnant.

%%%%%%%%%%%%%%%%%%%%%%%% begin of Fig.5 %%%%%%%%%%%%%%%%%%%%%%%%%%%
\vskip 0.3in
\begin{figure}
\centerline{
\psfig{figure=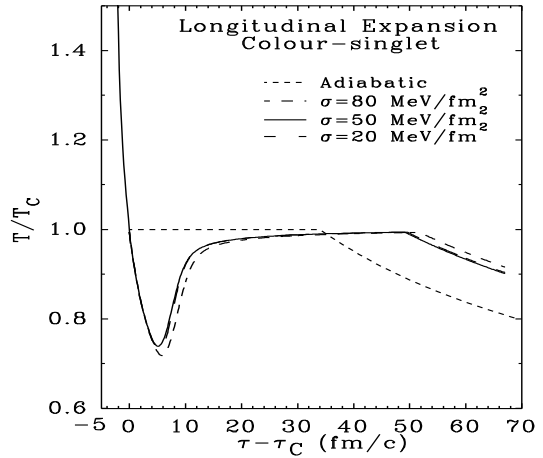,height=6cm,width=7cm}}
\vskip 0.3in
\caption{Same as Fig. 2 with various $\sigma$ values for colour singlet
case.}
\end{figure}
\vskip 0.3in
%%%%%%%%%%%%%%%%%%%%%%%%end of Fig.5 %%%%%%%%%%%%%%%%%%%%%%%%%%%

In Fig. 5 we have attempted to study the dynamics of phase transititon
as a function of quark/hadron interface tension $\sigma$. Recall~\cite{joe1}
that the dynamics depends sensitively on $\sigma$ when we do not impose
the restriction of colour singletness. With imposition of colour
singletness large variation in $\sigma$ leaves the dynamics fairly
unchanged. The reason for this is not too difficult to identify. Note that 
Eq.(\ref{ddf})
tells us that even if $\sigma =0$, there is possibility of nucleation of
hadronic bubble with a critical radius, ${R_b}^\star =\left [
3T/\pi(B-a_{qh})\right ]^{1/3}$. However as the nucleation rate, $I$, itself
depends on $\sigma$ via its prefactor, $I_0$ and $I=0$ if $\sigma=0$.
Thus we have kept $\sigma=50$ MeV/fm$^2$ as $I$ involves a more
realistic prefactor characterized by $\sigma$.

How will these findings differ for a more realistic (3+1) dimensional
expansion of plasma? Normally, the plasma is expected to expand mostly
in longitudinal direction initially. After a time $\tau 
\simeq R/c_{s}$ where R is transverse 
radius and $c_s$ is the speed of sound, the system is likely to expand 
in transverse direction as well. We, like the authors of Refs.~\cite{joe1},
take the other extreme and look on spherical expansion with the same initial
conditions as in Ref.~\cite{joe1}.
%%%%%%%%%%%%%%%%%%%%%%%% begin of Fig.6 %%%%%%%%%%%%%%%%%%%%%%%%%%%
\vskip 0.3in
\begin{figure}
\centerline{
\psfig{figure=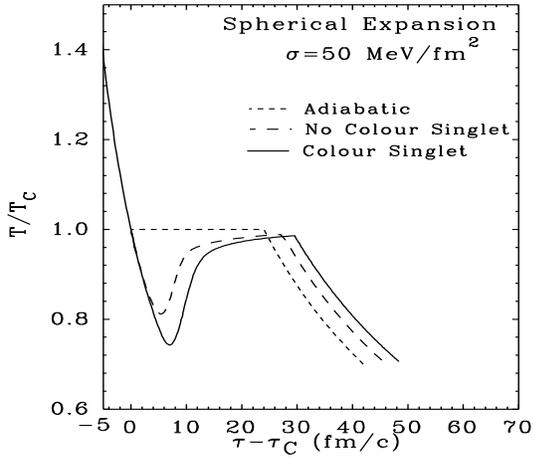,height=6cm,width=7cm}}
\vskip 0.3in
\caption{Same as Fig. 2 for matter undergoing spherical expansion.}
\end{figure}
\vskip 0.3in
%%%%%%%%%%%%%%%%%%%%%%%%end of Fig.6 %%%%%%%%%%%%%%%%%%%%%%%%%%%
%%%%%%%%%%%%%%%%%%%%%%%% begin of Fig.7 %%%%%%%%%%%%%%%%%%%%%%%%%%%
\begin{figure}
\centerline{
\psfig{figure=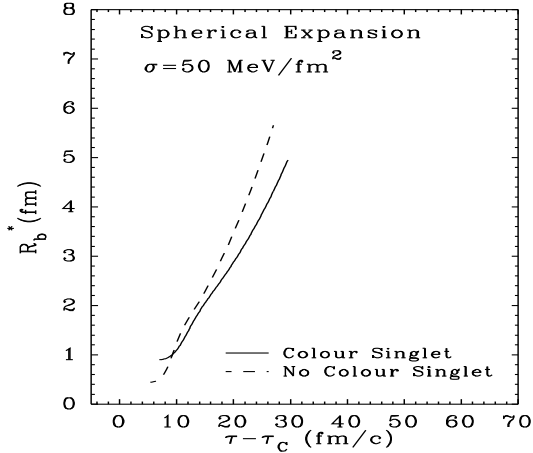,height=6cm,width=7cm}}
\vskip 0.3in
\caption{Same as Fig. 3 for matter undergoing spherical expansion.}
\end{figure}
\vskip 0.3in
%%%%%%%%%%%%%%%%%%%%%%%%end of Fig.7 %%%%%%%%%%%%%%%%%%%%%%%%%%%
Fig.6 is similar to Fig.2 with the exception that it considers spherical
(3-dimensional) expansion instead of longitudinal (1-dimensional) 
expansion. The degree of supercooling is almost of the order of 
longitudinal one ($30 \%$) with the
prime difference that hadronization is faster in 3-dimensional case.
This implies that the system spends less time in the neighbourhood of
$T_C$ than otherwise. Fig. 7 shows the variation of critical radius of
hadronic bubbles undergoing spherical expansion as a function of proper
time with and without colour singletness. It is clear that the critical
radius of nucleated hadronic bubbles for spherical expansion is smaller
compared to longitidunal expansion. In Fig. 8 we show the variation of
volume fraction converted to hadronic matter for spherical expansion as
a function of proper time. This is also clear from this figure that the
phase transition is faster than the longitudinal one. For the shake
of completeness we also present Fig.9 for spherical expansion which
shows the variation of temperature as a function of proper time for
different quark/hadron interface values. 

%%%%%%%%%%%%%%%%%%%%%%%% begin of Fig.8 %%%%%%%%%%%%%%%%%%%%%%%%%%%
\vskip 0.3in
\begin{figure}
\centerline{
\psfig{figure=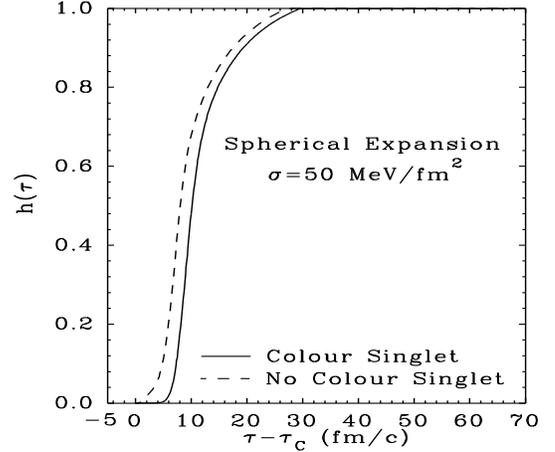,height=6cm,width=7cm}}
\vskip 0.3in
\caption{Same as Fig. 4 for matter undergoing spherical expansion.}
\end{figure}
\vskip 0.3in
%%%%%%%%%%%%%%%%%%%%%%%%end of Fig.8 %%%%%%%%%%%%%%%%%%%%%%%%%%%
%%%%%%%%%%%%%%%%%%%%%%%% begin of Fig.9 %%%%%%%%%%%%%%%%%%%%%%%%%%%
\begin{figure}
\centerline{
\psfig{figure=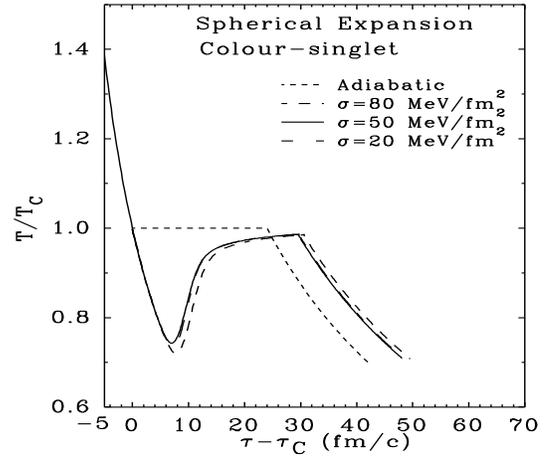,height=6cm,width=7cm}}
\vskip 0.3in
\caption{Same as Fig. 5 for matter undergoing spherical expansion.}
\end{figure}
\vskip 0.3in
%%%%%%%%%%%%%%%%%%%%%%%%end of Fig.4 %%%%%%%%%%%%%%%%%%%%%%%%%%%
\section{Nucleation Rate of Droplets of Quark-Gluon Plasma in Hot
Hadron Gas}
\label{drop}

The creation of QGP at AGS energies may proceed through the nucleation
of a plasma droplet in hot hadronic gas. Here the fields in plasma obey
the bag boundary conditions, staying inside the plasma droplet~\cite{mar}. 
If the radius of the droplet is $R_q$, the nucleation process is, 
generally, activated by the change in free energy which can be written 
within the bag model~\cite{mus,mar} as 
\bea 
\triangle F & =& T\ln \left ( \pi \sqrt 3\right ) + 4T\ln \left ( 
{8\over 3} V_q T^3 \right ) - a_q V_qT^4 \nonumber \\
&+& \left (B +P_h\right ) V_q +4\pi {R_q}^2 \sigma \ . \label{dfd}
\eea
Here, $V_q$ is volume of the plasma droplet formed in a superheated 
hadron gas. 
Usually, the nucleation rate of a plasma droplet from superheated
hadronic phase is estimated from Eq.(\ref{nra}). Now, the 
prefactor~\cite{joe2} is given by
\be
I_0 \ = \ {\kappa\over {2\pi}}\Omega_0 \ \ . \label{dpr}
\ee

The dynamical prefactor $\kappa$, determines the exponential 
growth rate of critical size droplets, and is given by~\cite{joe3}
\be
\kappa \ = \ {2\sigma\over {(\triangle w)^2 {{R_q}^\star}^3}} \ \left [ 
\lambda T + 2 \left ( {4\over 3} \eta +\zeta\right ) \right ] \ , 
\label{kap}
\ee
where  $\lambda$ is the thermal conductivity and $\eta$ and $\zeta$
are the viscosities of the hadronic phase. The bulk viscosity $\zeta$
is very small compared to shear viscosity $\eta$ and can be neglected. 
For these dissipative coefficients we use the parameterization 
of Danielewicz~\cite{dan}. ${R_q}^\star$ is the critical radius of a
nucleated plasma droplet. This can be obtained by minimizing 
$\triangle F$, given in Eq.(\ref{dfd}), with respect to $R_q$. 
To a first approximation the statistical prefactor~\cite{joe2,joe3} 
is given by
\be
\Omega_0 \ = \ {2\over {3\sqrt 3}} \left ( {\sigma\over T}\right )^{3/2} 
\left ( {{R_q}^\star \over \xi_h}\right )^4 \ . \label{om}
\ee
For our purpose we use following set of parameter $\sigma = 50$ MeV/fm$^2$,
$\xi_h = 0.7$ fm and $B^{1/4} = 200 $ MeV which gives $T_C\sim 170$ MeV.
We shall also give results for the prefactor $\sim T^4$ which is used
often in such studies.

As a first step, in Fig.~10 we plot the variation of $\triangle F$ as 
a function of
droplet radius with and without colour singlet restriction for three
different temperatures (160, 180, 240 MeV). Due to colour singlet
restriction, the height of $\triangle F$ is enhanced significantly in
each case. Fig.11 shows the variation of critical
radius of nucleated plasma droplet as a function of temperature. 
We see that the imposition of the colour singletness increases the
critical radius by factor of $\sim$ 3--5 over the range of temperatures 
that we have considered here. Knowing that creating a bigger size bubble 
by statistical fluctuation is considerably less probable, it is not 
surprising that this should lead to considerable suppression of nucleation 
rate when the degree of superheating is small. 

%%%%%%%%%%%%%%%%%%%%%%%% begin of Fig.10 %%%%%%%%%%%%%%%%%%%%%%%%%%%
\vskip 0.3in
\begin{figure}
\centerline{
\psfig{figure=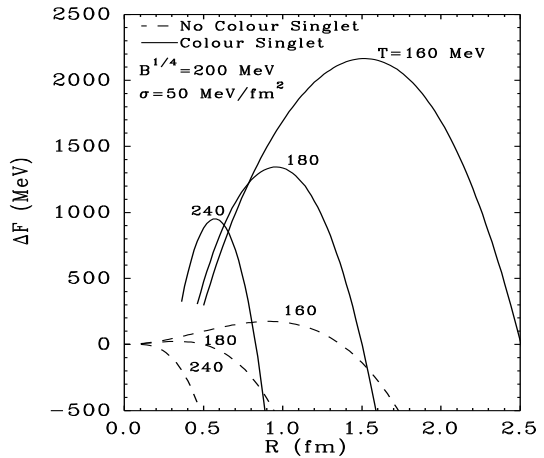,height=6cm,width=7cm}}
\vskip 0.3in
\caption{The free energy difference $\triangle F (R)$ for creation of a
plasma droplet in hot hadronic matter.}
\end{figure}
\vskip 0.3in
%%%%%%%%%%%%%%%%%%%%%%%%end of Fig.10 %%%%%%%%%%%%%%%%%%%%%%%%%%%
%%%%%%%%%%%%%%%%%%%%%%%% begin of Fig.11 %%%%%%%%%%%%%%%%%%%%%%%%%%%
\begin{figure}
\centerline{
\psfig{figure=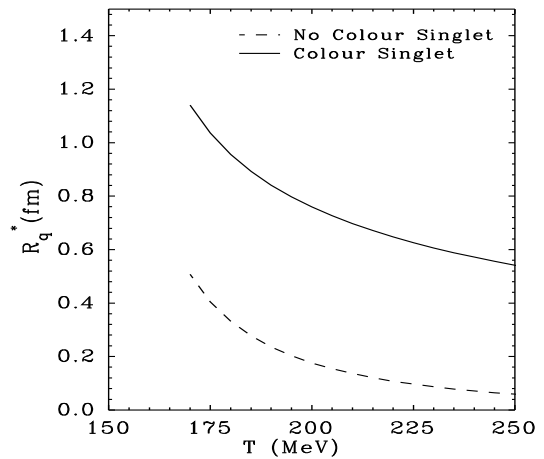,height=6cm,width=7cm}}
\vskip 0.3in
\caption{Critical radius of nucleated plasma droplets as a function of 
temperature in a superheated hadronic matter.}
\end{figure}
\vskip 0.3in
%%%%%%%%%%%%%%%%%%%%%%%%end of Fig.11 %%%%%%%%%%%%%%%%%%%%%%%%%%%
Fig.12 shows the variation 
of nucleation rate of plasma droplets in superheated hadron gas as a 
function of temperature. It is seen that the nucleation rate with the
prefactor $T^4$ (dashed lines) is suppressed considerably when the
restriction of colour singletness is imposed. Similar finding has been
reported recently in the literature~\cite{mad} with a somewhat different
expression for $\triangle F$. The results (solid lines) with the more 
realistic
prefactor (Eq.(\ref{dpr})) are richer in detail. Firstly we notice that
without the restriction of colour singletness the traditional prefactor
considerably overestimates the rate of nucleation. With prefactor given
in Eq.(\ref{dpr}), we see that switching on the requirement of colour
singletness lowers the rate of nucleation at smaller $T$. However, in a
surprising finding, we note that while the no-colour-singlet rate
decreases with increase of $T$, that with colour singlet restriction
increases and ultimately becomes larger at really high $T$.
This interesting behaviour is seen to  emerge from the
structure of the prefactor (Eq.(\ref{dpr})) which in fact is
proportional to ${R_q}^\star$, and ${R_q}^\star$ is seen to decrease
with increase in $T$ (Fig.11). We add here that Kapusta and 
Vischer~\cite{jv} have
recently proposed a more efficient mechanism for nucleation of plasma
droplet at AGS energies which envisages seeding of the plasma via
collision of two very energetic nucleons in the hot and dense hadronic
matter.
%%%%%%%%%%%%%%%%%%%%%%%% begin of Fig.12 %%%%%%%%%%%%%%%%%%%%%%%%%%%
\vskip 0.3in
\begin{figure}
\centerline{
\psfig{figure=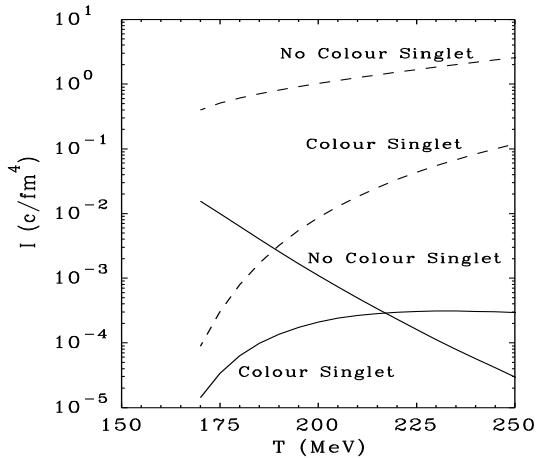,height=6cm,width=7cm}}
\vskip 0.3in
\caption{The nucleation rate of plasma droplets as a function of
temperature, in a superheated hadronic matter. The solid lines are 
with dynamical prefactor of Ref.~{\protect\cite{joe2}}
whereas those with dashed lines are with $T^4$ as prefactor.}
\end{figure}
\vskip 0.3in
%%%%%%%%%%%%%%%%%%%%%%%% begin of Fig.12 %%%%%%%%%%%%%%%%%%%%%%%%%%%

\section{Summary}
\label{con}

Before summarizing let us briefly examine some of the inputs
in the present work. While writing the expressions for the gain/loss
in free energy ($\triangle F$) we have included only the volume and 
the surface terms. 
This should be valid when the size of the plasma droplets or hadronic
bubbles is large. The prefacor we have used is valid for the situation 
when the size of the bubbles/droplets is larger than the correlation
length of the system ($\xi=0.7$ fm). We have seen that these 
bubbles/droplets have radii larger than a fermi, and thus both these
conditions are reasonably satisfies.

Recall that the surface energy coefficient $\sigma$ is zero in the
MIT bag model for massless quarks~\cite{fj}. Quarks traversing a hot
medium, do acquire a thermal mass. Finite temperature lattice
QCD~\cite{kkr} and a pure SU(N) gauge theory~\cite{bgkp} yield a 
value in the range $\sigma \approx$ 20--70 MeV/fm$^2$. We have 
thus mostly used $\sigma \approx$ 50 MeV/fm$^2$ and the effect 
of varying $\sigma$ is also studied. We have not come across any
estimate of the so-called curvature term in this case. We have 
verified, however, if we add a term for curvature energy in the 
expression of $\triangle F$ for the creation of hadronic bubble
in the plasma as in Ref.~\cite{mar}, then the supercooling of the
plasma is completely eleminated, if we ignore the colour singletness.
We have also checked that supercooling of plasma is possible if the
colour singlet restriction is imposed along with curvature 
contribution. However, for plasma droplet in hot hadronic gas 
the degree of superheating will be enhanced than otherwise.  

In brief, the theory, proposed recently~\cite{joe}, to describe the 
dynamics of hadronization has been generalized to
study the consequences of the requirement of colour singletness of QGP
which may be produced in relativistic heavy ion collisions. It is shown
that hadronization of longitudinally and spherically expanding plasma
may be slowed down due to this requirement. 
While there is no production of entropy for an adiabatic
phase transition, the  entropy increases by about $30\%$ in the treatment
of Ref.~\cite{joe}. The requirement of colour-singletness introduced
in the present work enhances this entropy by an additional $10\%$ and
also increases  the degree of supercooling.
We note an interesting possibility; a small fraction of 
QGP may not hadronize at all.
 
We also find that the nucleation of a plasma droplet in a superheated
hadronic matter is suppressed at low temperature ($T \le 220$ MeV),
but it is
enhanced at higher temperature ($T \ge 220$ MeV) when the colour singlet
restriction of QGP is accounted for. The decrease at $T\sim$ 170 MeV is
found to be by {\it four orders of magnutude} while the enhancement at 
250 MeV is by one order of magnitude.

It would be of interest to consider a generalization of this study to the
case of non-equilibrated plasma which hadronizes into a hadronic matter
having a richer equation of state. A (rather) crude result may be
obtained by keeping all other parameters like $T_C$, $\sigma$, $\eta$,
etc. fixed to their present values. This we feel, may not be quite
justified.
\vspace{0.6in}

We gratefully acknowledge useful discussions with Joe Kapusta during 
this work.

%\newpage


\begin{references}

\bibitem{klaus} K. Geiger, Phys. Rev. D {\bf 46}, 4965 (1992); {\it ibid.}
{\bf 46}, 4986 (1992); Phys. Rep. {\bf 258}, 237 (1995).
\bibitem{joe} L. P. Csernai and J. I. Kapusta, Phys. Rev. Lett. {\bf 69}, 
737 (1992). 
\bibitem{joe1} L. P. Csernai, J. I. Kapusta, Gy. Kluge, E. E. Zabrodin,
Z. Phys. C {\bf 58}, 453 (1993).
\bibitem{ll} L. D. Landau and E. M. Lifshitz, {\it Statistical Physics},
{\it 3rd ed.} (Pergaman, Oxford, 1980) Part I, Chap. 15
\bibitem{joe2} L. P. Csernai and J. I. Kapusta, Phys. Rev. D {\bf 46}
1379 (1992).
\bibitem{max} {\it see e.g.,} K. Kajantie, J. I. Kapusta, L. McLerran
and A. Mekjian, Phys. Rev. D {\bf 34}, 2746 (1986).
\bibitem{det} C. De Tar, Phys. Rev. D {\bf 32}, 276 (1985); C. De Tar and
J. Kogut, Phys. Rev. Lett. {\bf 59}, 399 (1987). 
\bibitem{red} K. Redlich and L. Turko, Z. Phys. C {\bf 5}, 201 (1980).
\bibitem{aub} G. Auberson {\it et  al.}, J. Math. Phys. {\bf 27}, 1658
(1986).
\bibitem{go} M. I. Gorenstein {\it et al.}, Phys. Lett. B {\bf 123}, 437 
(1983).
\bibitem{mus} M. G. Mustafa, Phys. Lett. B {\bf 318}, 517 (1994). 
\bibitem{mus1} M. G. Mustafa (under preparation).
\bibitem{ave} D. Amati and G. Veneziano, Phys. Lett. B{\bf 83}, 
87 (1979).
\bibitem{mw} G. Marchesini and B. R. Webber, Nucl. Phys. B{\bf 238},
1 (1984).
\bibitem{eg} J. Ellis and K. Geiger, Phys. Rev. D{\bf 54}, 949 (1996);
{\it ibid.} {\bf 54}, 1754 (1996).
\bibitem{mar} I. Mardor and B. Svetitsky, Phys. Rev. D {\bf 44}, 878 
(1991).
\bibitem{bura} L. Burakovsky, Phys. Lett. B{\bf 382}, 13 (1996).
\bibitem{mp} J. C. Miller and O. Panton, Phys. Rev. D {\bf 40}, 1789
(1989); {\it ibid.} {\bf 42}, 3334 (1990).
\bibitem{bj} J. D. Bjorken, Phys. Rev. D {\bf 27}, 140 (1983).
\bibitem{cf} F. Cooper, G. Frye and E. Sch${\ddot{\mathrm o}}$nberg,
Phys. Rev. D {\bf 11}, 192 (1974).
\bibitem{joe3} J. I. Kapusta, A. P. Vischer and R. Venugopalan, Phys.
Rev. C {\bf 51}, 901 (1995).
\bibitem{dan} P. Danielewicz, Phys. Lett. B {\bf 146}, 168 (1984).
\bibitem{mad} J. Madsen, D. M. Jensen and M. B. Christiansen, Phys. Rev. C 
{\bf 58}, 1883 (1996).
\bibitem{jv} J. I. Kapusta and A. P. Vischer, Phys. Rev. C {\bf 52}, 2725
(1995).
\bibitem{fj} E. Farhi and R. L. Jaffe, Phys. Rev. D{\bf 30}, 2379 (1984).
\bibitem{kkr} K. Kajantie, L. K\"arkk\"ainen and K. Rummukainen,
Nucl. Phys. B{\bf 333}, 100 (1990); S. Huang, J. Potvin, C. Rebbi and
S. Sanielevici, Phys. Rev. D{\bf 42}, 2864 (1990).
\bibitem{bgkp} T. Bhattacharya, A. Gocksck, C. Korthals Altes and 
R. D. Pisarski, Phys. Rev. Lett. {\bf 66}, 998 (1991).
\end{references}
\end{document}